\documentclass[usegraphicx,usenatbib]{mn2e}
\usepackage{amssymb}

\newcommand{\aap}{A\&A}                   
\newcommand{\apjs}{ApJS}                  
\newcommand{\apjl}{ApJ}                   

\newcommand{\erg}{\rm\thinspace erg}

\title[Quiescent X-ray variability in V404 Cyg]
{Characterizing the quiescent X-ray variability of the black hole low mass x-ray binary V404 Cyg}
\author[Bernardini \& Cackett]                                                    
{F.~Bernardini$^{1,2}$\thanks{E-mail: bernardini@wayne.edu}, 
E.~M.~Cackett$^{1}$\\
$^1$ Department of Physics \& Astronomy, Wayne State University, 666 W. Hancock St., Detroit, MI 48201, USA\\
$^2$ INAF, Osservatorio Astronomico di Capodimonte, Salita Moiariello 16, 80131 Napoli, Italia\\ }
\date{}

\def\Sw{{\em Swift}}
\def\XMM{{\em XMM-Newton}}
\def\CXO{{\em Chandra}}

\def\erg{\rm erg\,cm$^{-2}$\,s$^{-1}$}

\newcommand\beq{\begin{equation}}
\newcommand\eeq{\end{equation}}

\begin{document}
\label{firstpage}
\maketitle

\begin{abstract}

We conducted the first long-term (75 days) X-ray monitoring of the black hole low mass X-ray binary V404 Cyg, with the goal of understanding and characterizing its variability during quiescence. The X-ray light curve of V404 shows several flares on timescales of hours with a count rate change of a factor of about 5-8. The root mean square variability is F$_{\rm var}=57.0\pm3.2\%$. The first order structure function is consistent with both a power spectrum of index -1 (flicker noise), or with a power spectrum of index 0 (white noise), implying that the light curve is variable on timescales from days to months. The X-ray spectrum is well fitted by a power law with spectral index $\Gamma=2.10-2.35$, and we found that the spectral shape remains roughly constant as the flux changes. A constant spectral shape with respect to a change in the X-ray flux may favour a scenario in which the X-ray emission is dominated by synchrotron radiation produced in a jet.
 
\end{abstract}


\section{Introduction}

In low mass X-ray binaries (LMXBs) a compact object, a black hole (BH) or a neutron star (NS), is coupled with a late-type companion star, with mass lower than that of the Sun.Some LMXBs are persistent, constantly accreting matter from the companion through a standard accretion disk \citep{shakura73}. Consequently, they produce luminous X-ray emission (up to $10^{37-38}$ erg/s) close to the Eddington limit, $L_{\rm Edd}$. Other LMXBs are instead transient. They alternate long periods of low luminosity emission (quiescence), that can last years or decades,  to short and sporadic periods of intense activity (outbursts), lasting weeks-months. During quiescence the X-ray luminosity is of the order of $10^{31-33}$ erg/s. However, during outburst the X-ray luminosity dramatically increases by several orders of magnitude, and the sources become similar to persistent LMXBs. The quiescence to outburst cycle is broadly explained by the disk instability model  \citep[DIM, e.g.][]{cannizzo93,laso}, where the accreting matter is accumulated in the accretion disk during quiescence and then is suddenly transferred to the compact object during outburst. On the other hand, the origin of the X-ray emission from LMXBs during quiescence is still debated.  

Several models have been proposed to explain this quiescent state of low level of emission, like in the case of the family of models known as advection dominated accretion flow (ADAF) solutions, mainly developed by \cite{narayan3,narayan1,narayan2}, but see also \cite{ichimaru77} for some earlier idea. In this scenario, if the mass accretion rate from the companion drops below a critical value, and consequently the X-ray luminosity becomes very low, $\lesssim10^{-3}\,L_{\rm Edd}$, the accretion disk becomes unstable and likely undergoes a drastic structural change. The inner part of the disk, which in outburst extends close to the compact object, can evaporate, resulting in a disk truncated further out where the structure remains that of a standard disk (e.g. geometrically thin and optically thick). The evaporated gas instead assumes a spherical shape around the compact object and rotates much slower than the Keplerian velocity. The accretion timescale becomes more rapid than the cooling timescale, and thus the accretion flow is radiatively inefficient. The energy stored in the form of heat (entropy) in the gas is not radiated away (unlike what happens in a standard accretion disk) but is transported inward (advection).  In the case of a BH the energy is finally lost in the event horizon, and in the case of a NS it is transferred to the stellar surface. However, in order to explain the extremely low level of X-ray emission from some faint LMXBs, the action of an extra component in carrying away matter and radiation from the compact object is mandatory. The nature of this extra component is still a matter of investigation and debate, and likely depends on the characteristics of the specific accreting object. In the case of NSs, the accreting matter could be pushed away by the centrifugal force of the rotating magnetosphere \citep[see, e.g.][]{asai98, menou99}, or, alternatively, a `dead disk' could form, inhibiting but not completely preventing matter accreting onto the NS surface \citep[][]{dangelo10,dangelo12}. 
Other possibilities invoke the formation of gas outflows \citep[that the ADAF itself is inclined to produce, 
as first suggested by][]{narayan3,narayan1}, or a strong outflowing wind \citep[ADIOS;][]{blandford99} or a (radiatively inefficient) jet \citep[e.g.][]{fender03}, where the majority of the accretion power is released with the outgoing matter. 
Moreover, hybrid jet/ADAF models have been also suggested \citep[see, e.g.][]{yuan05}. See Sect. 3.6 of \cite{narayan08} for a clarifying review about ADAF plus outflow and jet models.  All the latter scenarios apply to both BH and NS LMXBs.

The X-ray spectrum of quiescent LMXBs shows no sign of the accretion disk (though the disk is X-ray bright during outburst and produces a bright thermal component then). For BHs, in particular, only a non thermal component is present. Its shape, between $\sim0.5$ and $\sim10$ keV, is a power law and its physical origin is still strongly debated. \cite{bildsten00} proposed that the quiescent X-ray emission of BH LMXBs could be generated by the stellar corona of the rapid rotating companion. However, \cite{lasota00} argued that the expected coronal emission would have likely been far below that observed in quiescent systems, as later unambiguously demonstrated by \cite{kong02}. Consequently, the X-ray emission of BH LMXBs, and its power law spectral shape, must be intrinsic to the accretion flow. 

The majority of LMXBs containing BHs, spend the bulk of their time in the quiescent state, this is also true for the nearby population of active galactic nuclei \citep[AGN, see e.g.][]{soria06,ho09,gallo10,pellegrini10}, or for the super massive BH in the center of our galaxy, Sgr A$^{*}$. However, also due to the low level of X-ray emission during quiescence, our knowledge of this state is still relatively poor.  The quiescence of several LMXBs, on the contrary to what "quiescence" might suggest, have been proven to be highly variable, and the origin of this variability is still unclear. For example, the NS LMXB Cen X-4 has shown X-ray variability on a wide range of timescale from hundreds of seconds up to years \citep{campana97, rutledge01, campana04, cackett10, cackett13, bernardini13}. The quiescent X-ray and the optical/UV emission of Cen X-4 are strongly correlated down to a timescale of a few hundreds seconds. A residual low level of accretion, down to the NS surface, that triggers optical/UV reprocessing on the companion star, must be the origin of this variability \citep{bernardini13}. Aql X-1, another NS LMXB, also showed X-ray variability on timescales from weeks to year \citep{rutledge02, cackett11,cotizelati13}. 

LMXBs containing BHs also seem to show some X-ray variability in quiescence on timescales of years \cite[e.g.][and more references therein]{reynolds13}. Moreover, V404 Cyg also showed a factor of 10--20 X-ray variability in less than a day, and a factor of $\sim2$ variability in $\sim30$ minutes \citep[][both work were made using a single X-ray observations lasting 16 and 60 ks only respectively]{wagner94,hynes04}.  However, up to now, regular X-ray monitoring of a quiescent BH over a long baseline (months) has not been performed. But, meaningful variability on such timescales is seen in persistent sources and those in outburst.  For instance, due to spectral state transitions \citep[see, e.g.][]{remillard06}, or other mechanisms such as precessing or warped disks \citep{charles08}. Motivated by this, and with the goal to shed some light on the quiescent state of BH LMXBs, we undertook a unique study monitoring the BH LMXB V404 Cyg with the \Sw\ satellite for approximately two months, on alternating days. This allowed us, for the first time ever for a BH LMXB in quiescence, to characterize the properties of this variability on timescales of days to months. We present the source in Sec. \ref{sec:v404}, the \Sw\ observations and data reduction in Sec. \ref{sec:obs}, the timing and spectral results in Sec. \ref{sec:results} that we later discuss in Sec. \ref{sec:discuss} (making also a comparison with the case of Cen X-4 in Sec. \ref{subsub:cenx4}). Finally, in Sec. \ref{sec:summary}, we report a brief summary of the main conclusions. 

\section{V404 Cyg} 
\label{sec:v404}

V404 Cyg, also known as GS 2023+338, was discovered in 1989 by the Ginga X-ray satellite during an outburst \citep{makino89}. While its distance was first believed to be about 3.5 kpc, more recent parallax observations, obtained with very long baseline interferometry at radio wavelengths, unambiguously put V404 Cyg at $2.40\pm0.14$ kpc \citep{millerjones09}. It is the most luminous of the quiescent BH LMXBs, with a X-ray luminosity of about $7\times10^{32}$ (d/2.4 kpc)$^{2}$ erg/s \citep{millerjones09}. It shows a transient behaviour with three confirmed outbursts: 1938, 1956, and 1989 \citep[see, e.g.][the latter for a detailed analysis of the 1989 outburst]{richter89,zycki99}. The companion is a $\sim1\;M_{\odot}$, K0($\pm1$) III--V star, almost filling its Roche lobe \citep{wagner92,casares93}. \cite{casares92} and \cite{casares94}, using spectroscopic optical data, measured an orbital modulation of $6.473\pm0.001$ d, and unambiguously derived the lower limit of the mass of the compact object $6.08\pm0.06\,{\rm M_{\odot}}$, that implies the presence of a black hole. By using infrared (IR) photometry, combined with spectroscopic data, \cite{shahbaz94} constrained the value of the mass ratio $q=16.7$ ($M_{x}/M_{c}$) and the inclination $i=56\pm4$ degrees. That led to an accurate measure of the black hole mass which is $12\pm2{\rm M_{\odot}}$. Later, \cite{shahbaz96} by using IR spectroscopy derived a $15\%$ upper limit to the contamination to the IR light by the accretion disk. This reduced the mass derived with IR photometry to $10\,{\rm M_{\odot}}$, still confirming that the compact object is a stellar mass black hole. During quiescence the source 0.5--10 keV X-ray spectrum is well fitted, as usual for quiescent BH LMXB, by a power law with spectral index $\Gamma\sim2$ multiplied by an interstellar hydrogen column density of $0.8-1.2\times10^{22}$ cm$^{-2}$ \citep{bradley07,plotkin13,reynolds13}. As it is the most luminous quiescent BH, and most of its main characteristics have been constrained, V404 Cyg is the best candidate to perform a study focused on X-ray variability during quiescence.    

\section{Observations and data reduction}
\label{sec:obs}

The \Sw\ satellite is a multi-wavelength instrument, with three telescopes on board that allow simultaneous coverage of the optical/ultraviolet band \citep[Ultraviolet Optical Telescope, UVOT][]{roming05,breeveld10}, the soft X-ray band (X-ray Telescope, XRT), and the hard X-ray band \citep[Burst Alert Telescope, BAT][]{gehrels04}. Because of the low X-ray flux of V404 Cyg, the source is undetected by BAT and, consequently, we only analysed XRT data, all taken in photon counting (PC) mode. Concerning the optical/UV analysis we explored data in the UVW1 filter (2600 \AA), 
the only one used. \Sw\ has observed the source for a total of 34 times with an exposure time of 5 ks each.
33 observations were performed, approximately every two days, as part of the main observational campaign, between 03 July 2012 and 16 September 2012 (obsid: 00031403002--00031403034). Moreover, an archival observation was performed on 26 April 2009 (obsid: 00031403001).

\subsection{XRT data reduction}

We used the web tool developed at the UK \Sw\ Science Data Center to produce the $0.3 - 10$ keV X-ray light curve, with the binning time corresponding to each pointing and each snapshot duration (the XRT pointings are always composed of a number of multiple snapshots of variable duration). This tool is available on the University of Leicester website\footnote{http://www.swift.ac.uk/} \citep[see][for a detailed description of the light curve extraction tool]{evans09}. The $0.3 - 10$ keV average source count rate is $0.014\pm0.008$ c/s, while the average background count rate is $0.002\pm0.001$ c/s. All uncertainties are hereafter at the $1\sigma$ confidence level. 

Due to the source faintness, in order to create spectra with the maximum level of signal to noise (S/N), we summed together spectra from different pointings with similar count rates (see Sect. \ref{subs:spec} for details about the selected pointings), using the following procedure. First, using \textsc{HEAsoft 6.12}, we generated the cleaned event file for each of the 34 pointings with the \textsc{xrtpipeline} command, together with the respective exposure maps, manually providing the best source position. Then, we summed the individual event files of interest by using \textsc{XSelect} and, one by one, the individual exposure maps by using the command $sum\,ima$ in the \textsc{XImage} package. We used a source extraction region of 20 pixels size, centered at the best source position, and a background extraction region of 60 pixels size, in a nearby clean region of the sky. Finally, we extracted the summed spectra from the summed event files with \textsc{XSelect} and generated the summed ancillary response file (ARF) using the summed exposure maps. Before fitting, all the spectra have been rebinned in order to have a minimum number of counts greater or equal to 20. This allows a correct use of the $\chi^2$ statistic. All spectral fits were performed with \textsc{XSpec 12.7.1} \citep{arnaud96} and the latest calibration files available in October 2012.

\subsection{UVOT data reduction} 

We performed the UVOT data reduction for each pointing by applying the four step procedure: 1) \textsc{uvotbadpix}, 2) \textsc{uvotexpmap}, 3) \textsc{uvotimsum}, 4) \textsc{uvotdetect}. We describe it in more detail in the following. From the raw image file, we generated the bad pixel map with \textsc{uvotbadpix}. Then, from the level II image fits file, we produced the exposure maps by using the previously generated bad pixel map and the \textit{uat.fits} file for the attitude file.  With the \textsc{uvotimsum} tool, we summed all the multiple image extensions contained in the level II image fits file of each pointing (the UVOT pointings are always composed of a number of multiple snapshots of variable duration). This procedure generates the total pointing image file.  With the \textsc{uvotimsum} tool we did the same for the exposure map, constructing the total exposure map of each pointing. Finally, we run the \textsc{uvotdetect} tool and we ascertained that V404 Cyg was not detected in any pointing in the UVW1 band.
 
In order to maximize the signal-to-noise ratio we summed together all available UVOT observations with the UVW1 filter. Then, we followed the above procedure, with the only difference that at step number 3 we added together images and exposure maps of different pointings.  We defined the source region with a circle of 5 arcsec size, centered at the best position for V404 Cyg, and the background region with a circle of 10 arcsec in a close by region free from other source contamination.  As a last step, using the \textsc{uvotsource} tool, we get a $3\sigma$ upper limit for the average source flux density in the UVW1 band of F$_{\lambda}<1.62\times10^{-18}$ erg s$^{-1}$ cm$^{-2}$ \AA$^{-1}$. This is consistent with the flux measured by \cite{hynes09} in the F250 Hubble Space Telescope band ($0.52\pm0.20\times10^{-18}$ \erg) which is close in wavelength to the \Sw\ UVW1 band.  

\section{Analysis and results}
\label{sec:results}

\subsection{X-ray light curve}

\begin{figure*}
\begin{center}
\includegraphics[angle=-90,width=11.4cm]{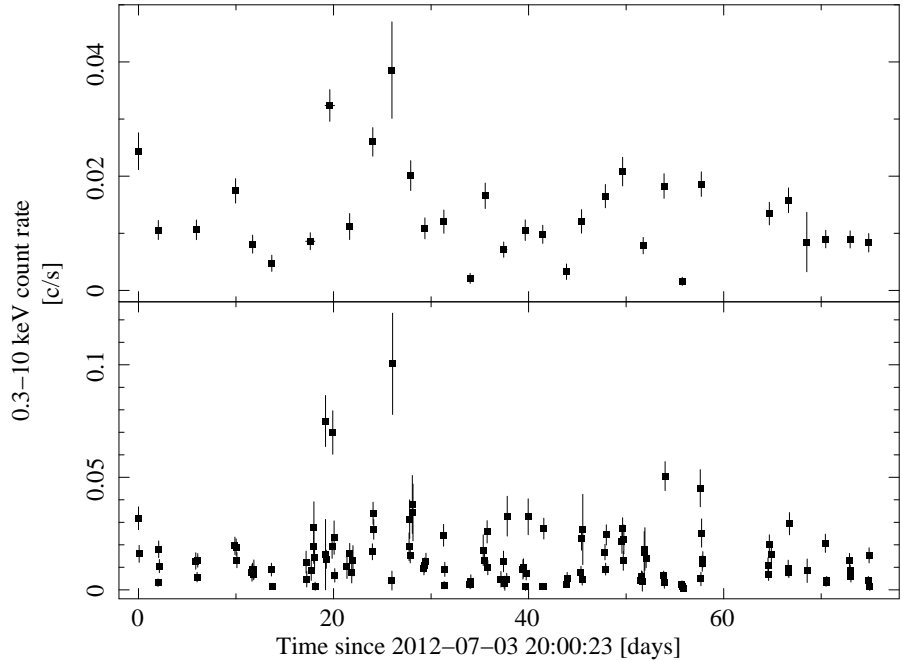} 
\caption{\textit{Top:} X-ray background subtracted light curve of V404 Cyg in the $0.3 - 10$ keV range on the XRT pointing timescale (average exposure time $\sim5$ ks). The reference time (2012-07-03 20:00:23) corresponds to the beginning of obsid 00031403002. The only point which is not plotted here corresponds to the archival obsid 00031403002 (t=-1166.25 days, $\sim0.15$ c/s). \textit{Bottom:} The same as the upper panel, but on the XRT snapshot timescale (average exposure time $\sim560$ s).}
\label{fig:lc}
\end{center}
\end{figure*}

We show the XRT 0.3--10 keV background subtracted light curve on the pointing timescale in Fig. \ref{fig:lc}. For plotting purposes we do not include the first (archival) pointing which does not belong to the main observational campaign (it is at $t=-1164.25$ d, and has a count rate of $0.015\pm0.002$ c/s).  The light curve is highly variable and shows dramatic count rate changes. The largest variability that is significant at greater than the $3\sigma$ level are changes of a factor of about 4--5. There are some larger changes that are significant at approximately $2\sigma$ level, like between 26.0 d ($0.039\pm0.008$ c/s) and 34.0 d ($0.0021\pm0.0009$ c/s), where we measure a decrease of a factor of $18.4\pm8.8$. In order to properly quantify the intensity of this variability we also calculated the root mean square variability \citep[see][for more details]{vaughan05}, finding F$_{\rm var}=57\pm3\%$. For comparison of the quiescent X-ray variability of V404~Cyg on different timescales, we also calculated F$_{\rm var}$ for an archival \XMM\ observation (obsid 0304000201) presented in \cite{bradley07}. Using the background-subtracted EPIC-pn lightcurve that is 34 ks long, we measure F$^{XMM}_{\rm var}=94\pm2\%$ when using a binning time of 1000 s.

We note that many of the \Sw\ light curve points lie well above or below $3\sigma$ from the average count rate of $0.014\pm0.008$ (see Fig. \ref{fig:lc}).  Moreover, significant rapid changes, between approximately the average level and higher or lower count rate, frequently happen on a two days timescale. See e.g. t$\sim18$ d, t$\sim20$ d, and t$\sim22$ d where the count rates is $0.009\pm0.002$ c/s,  $0.032\pm0.003$ c/s, and $0.011\pm0.002$ c/s respectively. These findings suggest that the source is variable both on a two-day timescale (the minimum distance between two pointings) and up to a seventy five days timescale (the maximum distance between two pointings).

We also produced the XRT $0.3 - 10$ keV background subtracted light curve on the snapshot timescale (average exposure time $\sim560$ s), which we show in the bottom panel of Fig. \ref{fig:lc}. Once again the light curve is highly variable, and thanks to the better time resolution, we clearly see the presence of multiple flares. The X-ray count rate increases and decreases, multiple times, by a factor of $\sim5-8$ on timescale of hours. See e.g. around 19 days where the count rate goes from $0.075\pm0.0011$ c/s to $0.0140\pm0.004$ c/s, implying a change of a factor $5.4\pm1.5$, or around 54 days where the count rate increases from $0.006\pm0.002$ c/s to $0.051\pm0.006$ c/s, implying a change of a factor $8.5\pm3.0$). However, we note that the lowest count rate bins for the snapshot timescale could correspond to statistical fluctuations below the mean flux at that time, and consequently there is the possibility that the amplitude of real variations is over-estimated. We cannot measure the root mean square variability for this snapshot lightcurve because the source is not detected in several of the snapshots due to the short exposures. 

\subsection{Structure Function}

The first order structure function V($\tau$) \citep[see][and references therein for more details]{do09} is a useful tool to study the variability in the time domain when the available dataset is unevenly sampled. V($\tau$) is related to the autocorrelation function and the power spectrum and represents another, straightforward, way to define the power of the variability at a given timescale. We computed  V($\tau$) only for the light curve in Fig. \ref{fig:lc} corresponding to the pointing timescale (it cannot be computed on the snapshot timescale light curve because of the multiple non detections). The first order structure function is defined as:
\begin{equation}
V(\tau) = \langle[s(t+\tau)-s(t)]^{2}\rangle
\end{equation}
where the angled bracket implies an average quantity, $s(t)$ is a set of measures (the count rate in the X-ray band) performed at the time $t$. We measured $[s(t+\tau)-s(t)]^{2}$ for all the possible pairs of time lags since our data set is unevenly sampled. We rebinned the time lags with a variable bin size to have a minimum number of pairs equal to 20 and a consistent number of pairs within each bin.  We take central bin value as the bin time lag, while the average of $V(\tau)$ is the value of the structure function for each bin at that time lag. The standard deviation of $V(\tau)$, divided by the square root of the number of points in the bin ($\sigma_{bin}/\sqrt{N_{bin}}$), represents the uncertainty in $V(\tau)$.

\begin{figure}
\begin{center}
\includegraphics[angle=-90,width=8.0cm]{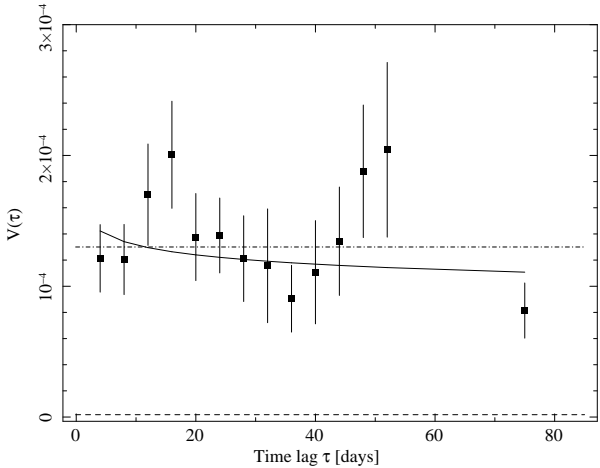} 
\caption{First order structure function V($\tau$) for the X-ray light curve (pointing timescale) of V404 Cyg in Fig. \ref{fig:lc} (upper panel). The dashed line corresponds to the intensity of the measurement errors (2$\sigma^2_{noise}\sim2\times10^{-6}$). The dot-dashed line corresponds to the expected intensity of the long time lag plateau ($2\sigma^2_{fluctuation}$). The solid line represents the fit made with a power law.}
\label{fig:struct}
\end{center}
\end{figure}

The expected shape of an ideal structure function consists of two plateaus connected by a curve whose slope is linked to the nature of the source variability (red noise, flicker noise etc.). The first plateau is connected to size of the measurement noise, and it has an amplitude of twice the variance of the measurement noise ($2\sigma^2_{noise}$). The second plateau usually occurs at a timescale greater than the inspected physical process, and it has an amplitude twice the variance of the fluctuations ($2\sigma^2_{fluctuation}$). The time interval where the curve connecting the two plateaus extends corresponds to the time scale where correlated variability is detected. The slope of this curve is linked to the slope of the power spectrum.  For example, the first order structure function of a power spectrum with $P(f) \propto f^{-2}$ (red noise) will be $V(\tau) \propto \tau^{1}$ \citep[see, e.g.][for more details and examples]{hughes92} in the region connecting the two plateaus. We show the first order structure function of V404 Cyg in Fig. \ref{fig:struct}. 

We fit the structure function with a power law of the form $y(t)=cx^{\gamma}$, finding $\gamma=-0.1\pm0.1$, $c=1.6\pm0.4\times10^{-4}$ with $\chi^2_{\nu}\sim1.1$ for 12 dof. Hence, the slope of the power law is consistent to be zero within $\sim1\sigma$, and thus $V(\tau) \propto \tau^{0}$. By looking at Fig. \ref{fig:struct}, we see that V($\tau$) has an intensity close to the value expected for the long time-lag plateau (2$\sigma^2_{fluctuation}\sim1.3\times10^{-4}$). Such a kind of plateau can be produced in different ways. A flat structure function could be produced by a flicker noise power spectrum, $P(f) \propto f^{-1}$, at a time lag where the fluctuations are still correlated, or it could also arise from a white noise power spectrum, $P(f) \propto f^{0}$, at a time lag where there is no more correlation. 

We confirm this by performing simulations of lightcurves with different power spectral slopes and determining their structure function. We simulated the limit of a dataset with no noise and much better sampling than the real case. If the two structure functions look the same for a high S/N dataset with better sampling, we cannot expect to detect a difference in the real data. We generated $10^{4}$ light curves with mean and standard deviation equal to that observed (0.140 c/s and 0.008 c/s respectively), for the cases of both a $f^{-1}$ and a $f^{0}$ power spectrum. The lightcurves were generated using the  \cite{timmer95} algorithm to produce the light curves that are 750 days long (10 times more than our data set), and sampled 10 times per day. Then, for each simulated light curve, we computed V($\tau$) by using  a binning that is three times smaller than that of Fig. \ref{fig:struct}. We fitted each simulated V($\tau$) with a power law function. We measure the slope $\gamma$ and the coefficient ($c$) of the power law and we derived the mean and the standard deviation of the distributions of the two values.  We get $\langle\,\gamma\,\rangle=0.11\pm0.11$, $\langle\,c\,\rangle=6.7\pm2.3\times10^{-5}$ for $f^{-1}$, and $\langle\,\gamma\,\rangle=-0.004\pm0.052$, $\langle\,c\,\rangle=1.1\pm0.2\times10^{-4}$ for $f^{0}$. Here the angled brackets imply the mean value of the distribution. We get fully consistent values for the two structure functions within uncertainty. We therefore conclude that both power spectra are producing two fully consistent structure functions with zero slope.  Summarizing, we also conclude that V404 Cyg is variable over the inspected timescale (2-75 days), but from the structure function we cannot determine if this variability is generated by a correlated (flicker noise) or an uncorrelated process (\textit{white}) noise.

\subsection{X-ray Spectral analysis}
\label{subs:spec}

Due to the source faintness, in order to search for spectral variability as a function of the X-ray flux, we selected three count rate ranges: low $<0.01$ c/s, medium 0.011--0.020 c/s, and high $>0.020$ c/s (see Fig. \ref{fig:lc}), and we produced three summed spectra\footnote{The three spectra are made with the sum of the following obsid $0003140300+$: 2,9,11,12,13,24 (high); 1,5,15,17,22,23,26,28,29,30 (medium), 3,4,5,7,8,10,14,16,18,20,21,25,27,31,32,33,34 (low)}. This gives roughly the same number of counts for each spectrum.  The high spectrum has $\sim465$ counts for $\sim23.7$ ks of exposure, the medium spectrum has $\sim550$ counts for $\sim41.9$ ks of exposure, and the low spectrum has $\sim490$ counts for $\sim74.1$ ks of exposure. The spectra cover the 0.3--10 keV range, and we are mainly interested in verifying the presence of spectral changes with the flux.  We first restricted our analysis to the simplest spectral model usually applied to BH LMXB in quiescence. It consists of a power law multiplied by the $phabs$ component, which accounts for the photoelectric absorption due to the interstellar medium.  We started fitting the three spectra simultaneously with all the components free to vary.  We verified that the value of $N_{H}$ was constant within uncertainties among the three.   Thus, we imposed $N_{H}$ to be the same between different spectra, but free to vary to minimize the $\chi^2$, and we determined a new fit.  We report the value of all model components, together with the 0.3--10 keV absorbed/unabsorbed flux, the 0.3--10 keV luminosity and the bolometric luminosity in Tab. \ref{tab:spec}. The latter was calculated by extrapolating our model to the 0.01--100 keV energy range, with the caveat that the true spectral shape is unknown both above 10 keV and much below 1 keV. 

\begin{figure}
\begin{center}
\includegraphics[angle=-90,width=8.4cm]{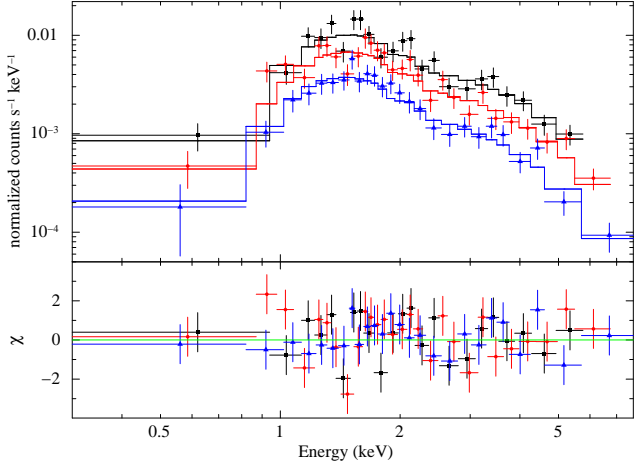} 
\caption{Summed spectra of V404 Cyg for three count rate ranges: low $<0.011$ c/s (blue triangles), medium 0.011--0.020 c/s (red circles), and high $>0.020$ c/s (black squares). The solid lines represent the model made by a power law multiplied by $phabs$ (photoelectric absorption). Residual are shown in the lower panel.}
\label{fig:spec}
\end{center}
\end{figure}

\begin{figure}
\begin{center}
\includegraphics[angle=-90,width=8.4cm]{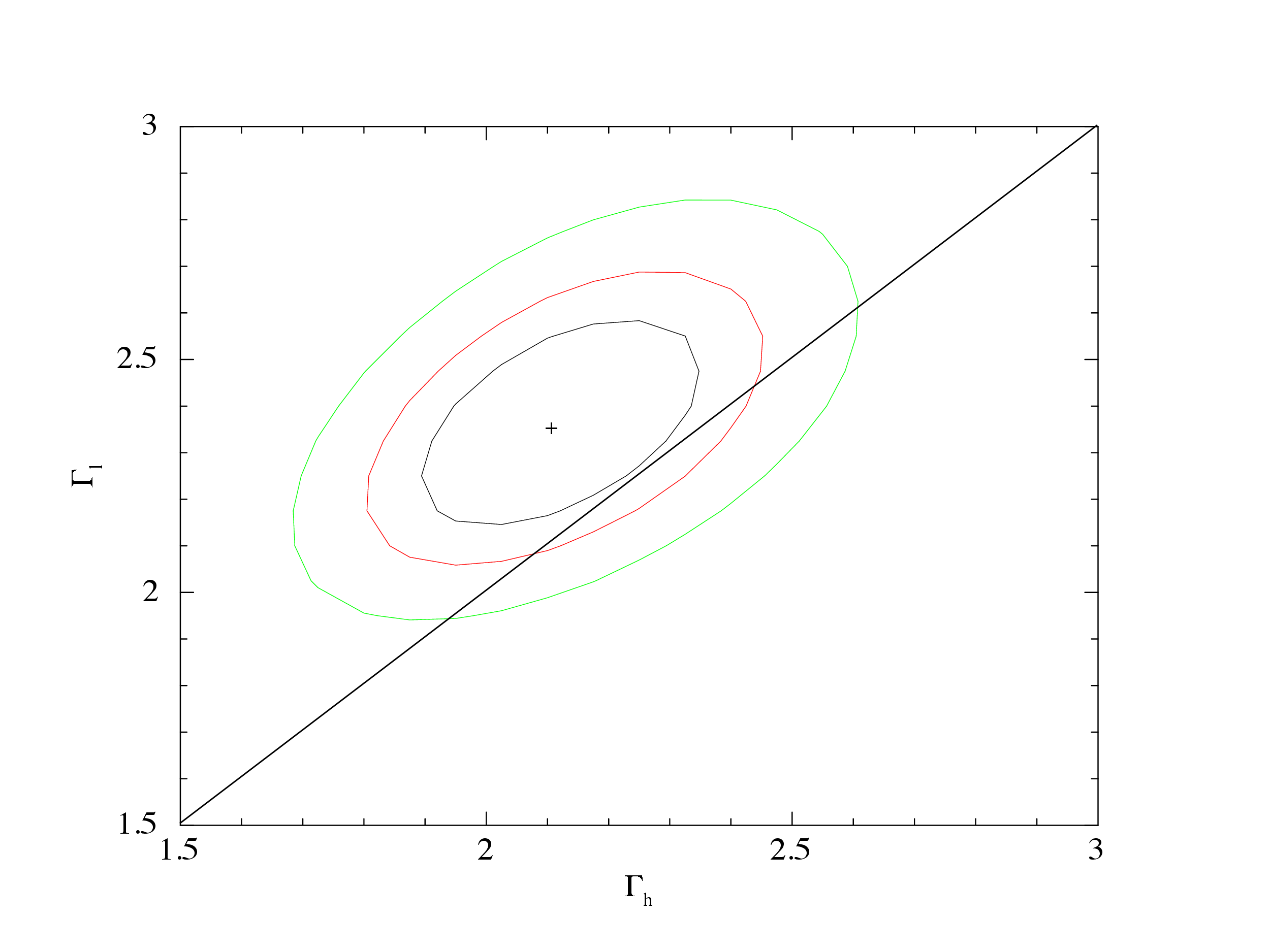} 
\caption{Contour plot showing the allowed region in the $\Gamma_{h}$-$\Gamma_{l}$ plane for our best fit power law model on the three count rate spectra. The three contours illustrate the $68\%$, $90\%$, and $99\%$ confidence level (black, red, green). The solid straight line represents $\Gamma_{h}=\Gamma_{l}$.}
\label{fig:contour}
\end{center}
\end{figure}

The spectral fits are shown in Fig. \ref{fig:spec}.  We measure $N_{H}=8.9\pm0.9\times10^{21}$ cm$^{-2}$, consistent with that previously reported for the source in quiescence \citep[e.g.][]{bradley07,plotkin13}, and slightly higher than the total in the direction of the source that lies between $6.6\times10^{21}$ cm$^{-2}$ and $8.1\times^{21}$ cm$^{-2}$ \citep[][respectively]{kalberla05,dickey90}. This could imply the presence of localized absorbing material intrinsic to the source \citep[as also suggested by e.g.][]{wagner94}.  We get a roughly constant power law index, $\Gamma_{l}=2.10\pm0.15$, $\Gamma_{l}=2.20\pm0.14$, $\Gamma_{l}=2.35\pm0.15$, consistent with that for the source in quiescence \cite[e.g.][who analysed \CXO\ and \XMM\ data respectively]{bradley07,plotkin13}. To further verify the consistence between $\Gamma_{l}$ and $\Gamma_{h}$, we produced a contour plot of the two components that we show in Fig. \ref{fig:contour}. The two power law indices are consistent within a little more than the $1\sigma$ confidence level. However, a small change, less than $11\%$ could be present, but not detectable with present data. We defined here the maximum percentage change, as the absolute value  of the difference between $\Gamma$ of the high spectrum and that of the low spectrum, divided by its average value, e.g.: 3($\mid\Gamma_{h}$-$\Gamma_{l}\mid$)/($\Gamma_{h}$+$\Gamma_{m}$+$\Gamma_{l}$). We conclude that, within statistical uncertainty, only the power law normalization is clearly changing. This suggests that, likely, the spectral shape remains almost the same as the flux changes.

\begin{table*}                                                                                                                                                                                                                                                                                                                                                                                                                                                   
\caption{Spectral results as a function of the count rate: low $<0.011$ c/s, medium 0.011--0.020 c/s, and high $>0.020$ c/s ($\chi^2_{\nu}=1.14$, 70 dof). The three spectra are fitted together imposing a common column density (free to vary) $N_{H}=8.9\pm0.8\times10^{21}$ cm$^{-2}$. The results of fit on the spectrum made summing together all the existing pointing (total) are also reported. For the latter we get $N_{H}=9.1\pm0.9\times10^{21}$ cm$^{-2}$ ($\chi^2_{\nu}=1.22$, 68 dof). The 0.3--10 keV absorbed flux, the 0.3--10 keV unabsorbed flux, and the 0.3-10 keV and 0.01--100 keV luminosity at 2.4 kpc are also reported. Uncertainty are at $1\sigma$ cl.}
\begin{center}
\begin{tabular}{cccccc}
\hline 
\\
Parameter                                    & Unit         & High                 & Medium             & Low           & Total           \\
\hline                                                                                                                              
$\Gamma$                                     &              & $2.10\pm0.15$        & $2.20\pm0.14$      & $2.35\pm0.15$ & $2.31\pm0.11$    \\
Norm.                                        & 10$^{-4}$    & $5.1\pm0.8$          & $3.5\pm0.5$        & $2.1\pm0.4$   & $3.1\pm0.4$     \\
\\                                                                                                                                     
F$^{absorb}_{0.3-10\, \rm keV}$               & 10$^{-12}$   & $1.20\pm0.10$          & $0.75\pm0.05$      & $0.37\pm0.03$ & $0.58\pm0.03$   \\
                                             & erg/s/cm$^2$ &                      &                    &               &                 \\\\
F$^{unabsorb}_{0.3-10\, \rm keV}$            & 10$^{-12}$   & $2.7\pm0.7$              & $1.7\pm0.5$            & $1.0\pm0.3$       & $1.5\pm0.3$       \\
                                             & erg/s/cm$^2$ &                      &                    &               &                 \\\\       
L$^{\rm 2.4\,kpc}_{0.3-10\, \rm keV}$ $^a$   & 10$^{33}$    & $\sim1.9$            & $\sim1.2$          & $\sim0.7$     & $\sim1.0$       \\
                                             & erg/s        &                      &                    &               &                 \\
L$^{\rm 2.4\,kpc}_{0.01-100\, \rm keV}$ $^a$ & 10$^{33}$    & $\sim5.5$            & $\sim4.1$          & $\sim3.2$     & $\sim4.3$       \\
                                             & erg/s        &                      &                    &               &                 \\\\
                                                                                                                                    
$\chi^2_{\nu}$ (dof)                         & \multicolumn{4}{c}{1.14  (70)}      &  1.22 (68)                                           \\ 
\hline 
\end{tabular}
\label{tab:spec}
\end{center}
\end{table*}

\begin{figure}
\begin{center}
\includegraphics[angle=-90,width=8.0cm]{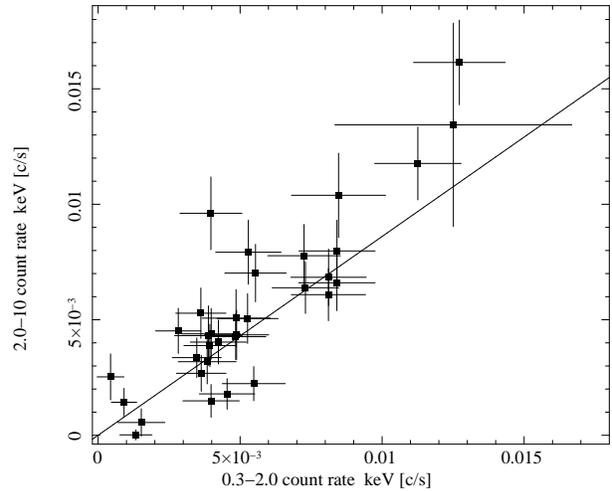} 
\caption{0.3--2.0 keV count rate vs 2--10 keV count rate. The solid line represents the best fit made with a linear component.}
\label{fig:fluxflux}
\end{center}
\end{figure}

To investigate how the spectrum changes on shorter timescales, we compared the 0.3--2 keV count rate with that in the 2--10 keV range for each XRT pointing (see Fig.~\ref{fig:fluxflux}). This allows us to explore the spectral variability between every observation (with two-day spacing) rather than just comparing three average spectra. We fit a straight line to the 0.3--2 keV versus 2--10 keV count rate relation,  which gives a best-fitting slope of $0.86\pm0.04$ with $\chi^2_{\nu}=3.6$. Although some scatter is present, making the $\chi^2_{\nu}$ value high, there is no significant deviation from a linear fit,  confirming that the overall spectral shape remains roughly constant as the flux changes.  The fact that the slope of the straight line is 0.83 simply implies that the flux in the 0.2--10 keV band is lower than that in the 0.3--2 keV band (i.e.  $F_{2-10 keV}=0.86\,F_{0.3-2 keV}$). However, the relative flux in the two bands stay the same as the source varies.  

\begin{figure}
\begin{center}
\includegraphics[angle=0,width=8.4cm]{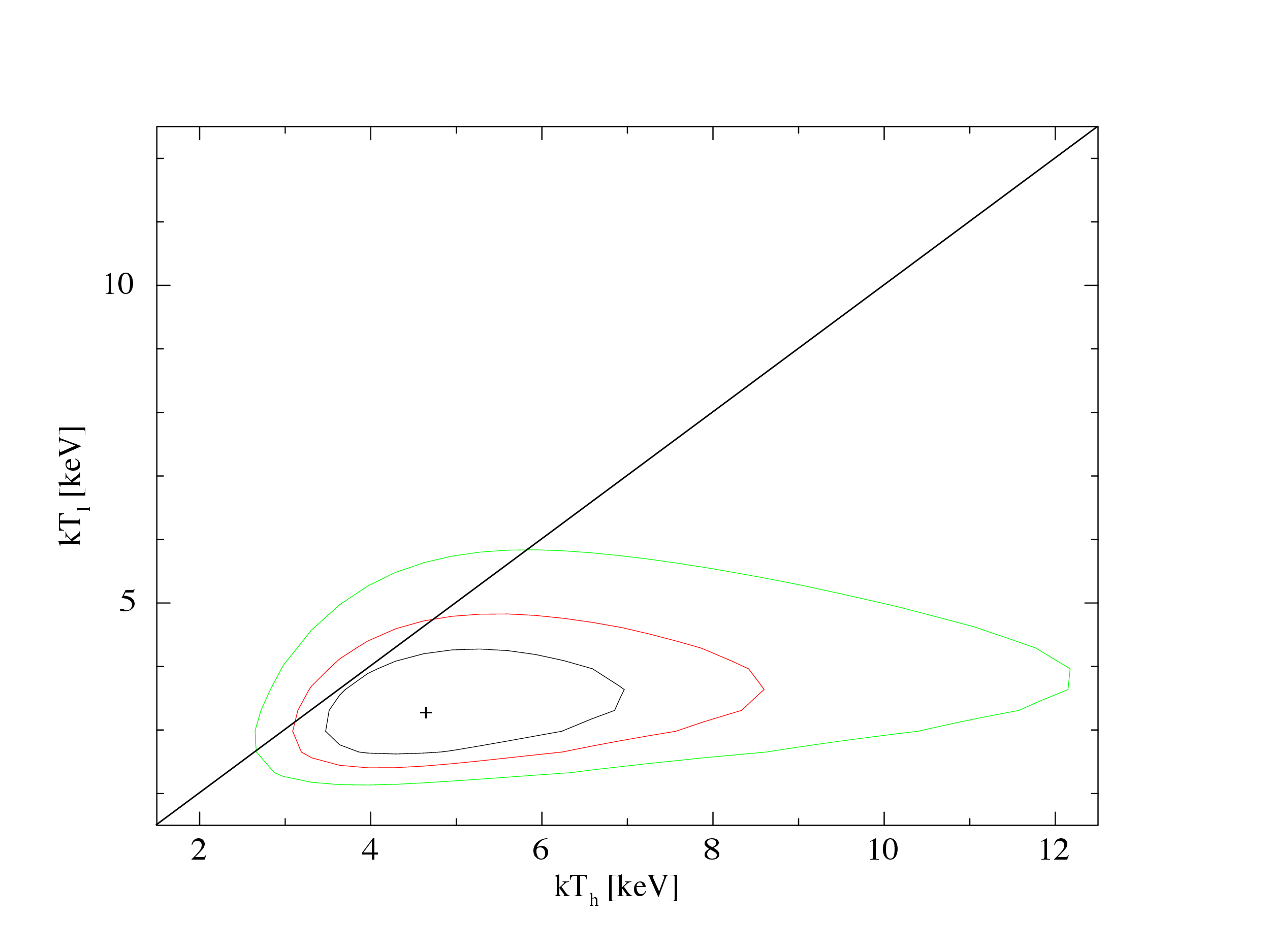} 
\caption{The same as Fig. \ref{fig:contour} but for the bremsstrahlung model, where $kT_{h}$ and $kT_{l}$ is the bremsstrahlung temperature for the high and low count rate spectrum respectively.}
\label{fig:contourkT}
\end{center}
\end{figure}

\begin{figure}
\begin{center}
\includegraphics[angle=0,width=8.4cm]{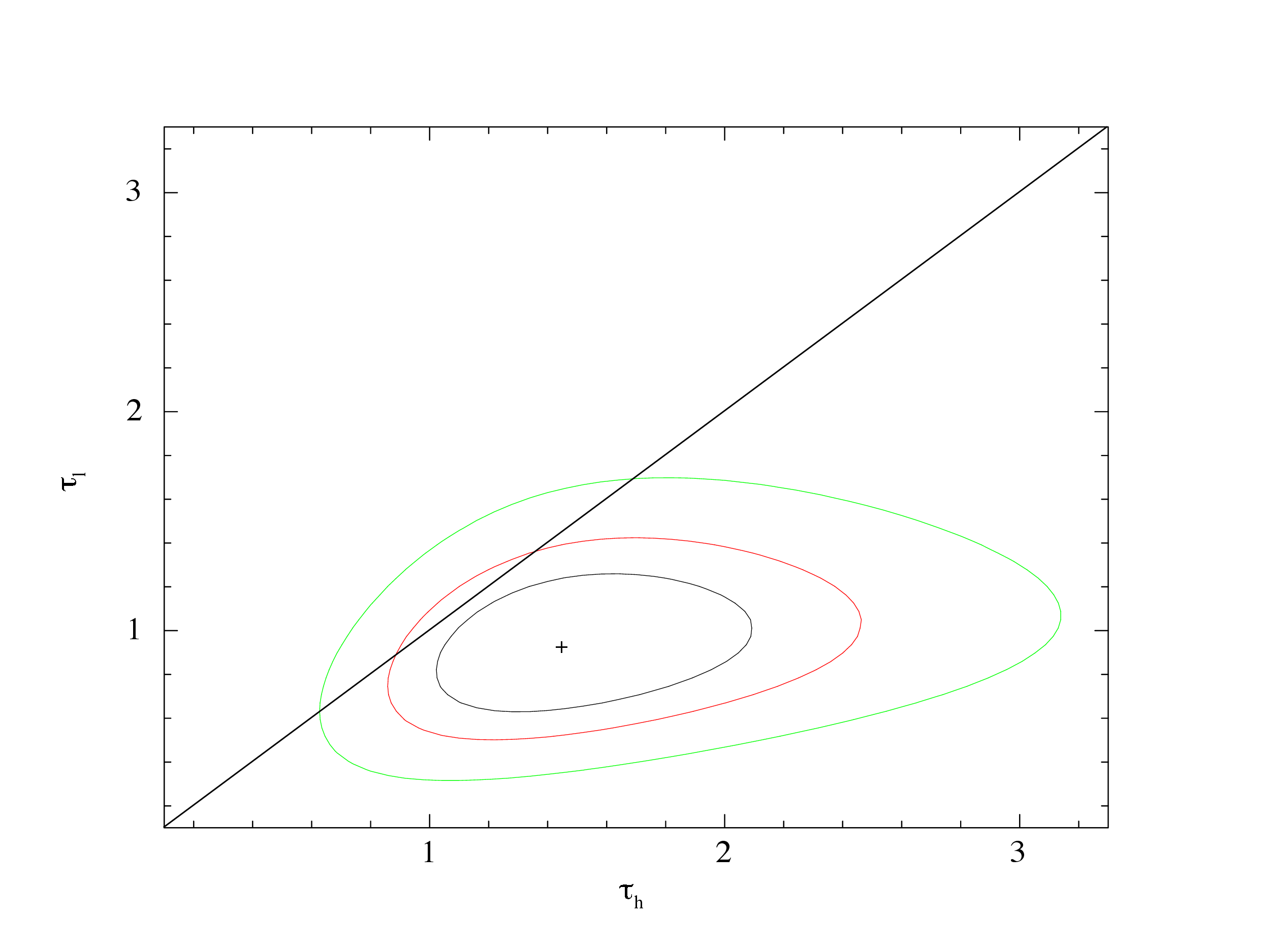} 
\caption{The same as Fig. \ref{fig:contour} but for the $CompTT$ model, where $\tau_{h}$ and $\tau_{l}$ is the plasma optical depth for the high and low count rate spectrum respectively.}
\label{fig:contourtau}
\end{center}
\end{figure}

As a second step we fitted the three spectra with two different more physical models that are normally used for BH LMXBS.  The first one is appropriate for the X-ray emission from hot electrons and ions in the outer part of an ADAF region, and consists of a bremsstrahlung component. The second one is appropriate X-ray emission arising from both a spherical Comptonizing corona or from the inner part of an ADAF region, and consists of a $compTT$ component \citep{tit94}. Both models are multiplied by $phabs$ which we imposed to be the same among different spectra, but free to vary.  We report all fitting results in Tab \ref{tab:spec2}.  Concerning the bremsstrahlung model, kT it is constrained to lie in the range 3.3--4.7 keV, and only the normalization clearly varies. We get $\chi^2_{\nu}=1.17$ for 70 dof. The inspection of the contour plot of kT$_{h}$ vs kT$_{l}$ shows that the two are consistent within $\sim1\sigma$ (see Fig. \ref{fig:contourkT}).  For the $compTT$ model, which has a higher number of free parameters, we first fixed the temperature of the Comptonizing corona to a reasonable value, kT=60 keV (this component is otherwise unconstrained due to the fact that we are covering here the 0.3--10 keV energy range only). Then, we verified that the temperature of the seed photons (T$_{0}$) was constant within uncertainties between different spectra. Consequently, we imposed it to be the same between different spectra, but free to vary, in order to minimize the $\chi^2$, and we tried to put some constrain on the variability of the plasma optical depth ($\tau_{p}$). We get T$_{0}=0.33\pm0.07$, and an optical depth, $\tau_{p}=0.9-1.5$, that is constant within slightly more than $1\sigma$ (see Fig. \ref{fig:contourtau}), and only the normalization clearly changes. In both cases, bremsstrahlung and $compTT$ models, the spectral shape remains roughly the same, within statistical uncertainty, as the flux varies. We note that the derived optical depth is high, but it depends on the value of kT, meaning that for increasing value of kT $\tau_{p}$ decreases. Moreover, we are interested in measuring the percentage variability range of $\tau_{p}$, more than is absolute value. Such modeling allows us to constrain the maximum change in the ion and electron temperature (kT$^{brem}$) and the optical depth of the Comptonizing corona ($\tau_{p}$) respectively.  We measure $\Delta\,KT^{brem}<34\%$ and $\Delta\tau_{p}<49\%$ respectively. The latter values are less constrained than that obtained with a power law model ($\Delta\Gamma<11\%$), where no assumptions are made (while for the $CompTT$ model we arbitrarily fixed the value of $T_0$). Consequently, in order to quantify the maximum fractional spectral change between different count rate spectra, we use the power law model.

\begin{table*}                                                                                                                                                                                                                                                                                                                                                                                                                                                   
\caption{Spectral results as a function of the count rate: low $<0.011$ c/s, 
medium 0.011--0.020 c/s, and high $>0.020$ c/s for the bremsstrahlung and the $compTT$ model.
Uncertainty are at $1\sigma$ cl.}
\begin{center}
\begin{tabular}{ccccc}
\hline 
\multicolumn{5}{c}{Bremsstrahlung}      \\ 
Parameter                                       & Unit                & High                        & Medium                & Low                    \\
\hline
N$_{\rm H}$                                  & 10$^{22}$ cm$^{-2}$ &    \multicolumn{3}{c}{$0.68\pm0.06^{a}$}       \\                                                                            
KT$^{brem}$                                  &   keV               & $4.7\pm^{1.3}_{0.9}$      & $4.2\pm^{0.9}_{0.7}$    & $3.3\pm^{0.6}_{0.5}$ \\
Norm.                                             & 10$^{-4}$      & $4.2\pm0.5$          & $2.8\pm0.3$        & $1.7\pm0.2$      \\

$\chi^2_{\nu}$ (dof)                         &                    &                                   & 1.17  (70)           & \\                                       
\hline 
\multicolumn{5}{c}{CompTT} \\
Parameter                                       & Unit                & High                        & Medium                & Low                    \\
\hline
N$_{\rm H}$                                  & 10$^{22}$ cm$^{-2}$ &    \multicolumn{3}{c}{$0.5\pm0.2^{a}$}         \\          
T$_{0}$                                          &  keV                           &    \multicolumn{3}{c}{$0.33\pm0.07^{a}$}      \\                  
$\tau_{p}$                                    &   keV                          & $1.5\pm0.4$      & $1.3\pm0.3$    & $0.9\pm0.2$ \\

Norm.                                           & 10$^{-6}$      & $9\pm2$          & $6\pm2$        & $3\pm1$      \\

$\chi^2_{\nu}$ (dof)                     & &&1.12  (69) &\\ 
\hline 
\end{tabular}
\label{tab:spec2}
\end{center}
$^{a}$ The value is imposed to be the same between different spectra, but is left free to vary to minimize the $\chi^2$.
\end{table*}

\subsubsection{Energetics}

We estimated the flux in the UVW1 band, by summing together all available pointings. The source is undetected in the UVW1 band, with an lower limit to its magnitude (AB system) $>25$.  The $3\sigma$ upper limit to the source flux density is $F_\lambda < 1.62\times10^{-18}$ \erg\AA$^{-1}$.  Comparing with Fig. 7 from \cite{hynes09} it can be seen that the \Sw\ UVW1 filter is covering a region of the source spectral energy distribution ($\sim1.1\times10^{15}$ Hz)  where the flux dramatically drops due to Galactic extinction.  To convert the flux density limit to a luminosity limit we first convert from equivalent hydrogen column to E(B-V) dereddening the flux using the gas-to-dust ratio, $N_H ({\rm cm^{-2}})=6.86\times10^{21}$ E(B-V) from \cite{guver09}. Finally, to estimate the extinction correction at 2600 \AA (UVW1) we use the interstellar extinction curve of \cite{cardelli89}, getting 3760 as conversion factor (this means that the dereddened flux is 3760 times the observed flux). We therefore obtain a $3\sigma$ upper limit to the UVW1 luminosity $L_{\lambda}<4.2\times10^{30}$ erg/s  ($\lambda\,L_{\lambda}<1.1\times10^{34}$ erg s$^{-1}$). 

We get an estimate of the X-ray luminosity by fitting the total spectrum (summed from all pointings), once again, with a model consisting of a power law multiplied by $phabs$. We report all best fit model parameter in Tab. \ref{tab:spec}. We infer a 0.3-10 keV unabsorbed flux of $1.5\pm0.3\times10^{-12}$ \erg. Assuming a distance of 2.4 kpc and a spectral power law shape (that is expected to extend above 10 keV) we extrapolate this to a 0.01--100 keV luminosity L$_{x}\sim4.3\times10^{33}$ erg/s. The latter and the following values should be taken as indicative only, since the real spectral shape both above 10 keV and much below 1 keV is not known. Since the spectral change is almost constant as the flux varies, we can get an estimate of the 0.01--100 keV luminosity corresponding to the two \Sw\ snapshots with the highest and lowest count rate respectively,  by simple proportionality. We get L$_{max}\sim2.8\times10^{34}$ erg/s (obsid 00031403012, $\sim0.1$ c/s) which implies $\sim2\times10^{-5}$ L$_{\rm Edd}$, and L$_{min}\sim1\times10^{33}$ erg/s (obsid 00031403027, $\sim0.002$ c/s) which implies $\sim8\times10^{-7}$ L$_{\rm Edd}$, where L$_{\rm Edd}=1.26\times10^{39}$ erg/s, for a 10 M$_{\odot}$ BH.

\section{Discussion}
\label{sec:discuss}

We showed that the quiescent X-ray light curve of V404 Cyg is highly variable on timescales of days to months, with the fractional root mean square variability $F_{\rm var}=57.0\pm3.2\%$. The first order structure function, $V_{\tau}$, has a shape consistent with both a flicker noise power spectrum (index $-1$) and a white noise power spectrum (index $-0$). This implies that variability is detected at all inspected timescales, from about 2 days up to about 75 days. However, the structure function analysis cannot determine if the variability is due to a correlated or to an uncorrelated process. We observed multiple times a factor of 4--5 variability on timescales of about a week. Higher magnitude changes (a factor of 10--20) have been previously observed on $\sim12$ hours timescale from V404 Cyg during quiescence by \cite{wagner94,hynes04}. We also showed that multiple flares are presents in the X-ray light curve on timescale of about one hour, with the count rate changing by a factor of $\sim5-8$. A factor of a few variability has been already seen on both $\sim30$ minutes and years timescales \citep{wagner94,reynolds13}. Consequently, V404 Cyg is variable during quiescence  on a wide range of timescales from tens of minutes to years.
 
The power law observed in the energy spectrum of BH LMXBs is likely generated in the inner region of the accretion flow, closer to the BH, where a population of high-energy electrons is present. Moreover, there is no sign of the presence of the accretion disk in the X-ray spectrum of V404 Cyg during quiescence (nor from quiescent BH LMXBs in general) which would instead produce thermal emission.  Consequently, the observed variability is not originating in the accretion disk, but is likely driven by mass accretion rate fluctuations in the inner region of the accretion flow, close to the BH. We can only speculate about the real nature of the matter in the inner accretion flow and about how it is triggering the variability.
 
If the matter around the BH is in an ADAF state, hot electrons in the outer region of the ADAF (at a distance $\gtrsim10^{4}$ Schwarzschild radii from the central BH) should cool down through thermal bremsstrahlung \citep[see, e.g.][]{narayan3,narayan1}. However, this emission should be consistent with a power law spectral shape with index 1, while here we observe a softer power law with index $2.10-2.35$.  We conclude, as also suggested by \cite{quataert99}, that the soft X-ray emission we observe is not thermal bremsstrahlung arising from the outer region of an ADAF, but it could be instead inverse Compton emission produced by hot electron close to the BH. However, \cite{hynes09} analysed the spectral energy distribution of V404 Cyg in quiescence and argued that the contribution to the accretion light in the UV band is minimal,  disfavouring an ADAF scenario alone, that instead should produce a more abundant UV emission. On the other hand, an ADAF plus an outflowing wind \citep[see, e.g.][]{quataert99} could be consistent with the SED of V404 Cyg (the wind has the effect to suppress the UV emission from the ADAF itself). We emphasize that a formal fit of this model against real data has never been performed so far. 

Hot electrons surrounding the black hole, and emitting a Compton like spectrum, are also part of the so-called jet models \citep[see, e.g.][and more reference therein]{gallo07}.  In this scenario, inverse Compton emission is produced at the base of the jet, in a region linked to a compact and magnetic corona, where there is a population of quasi-thermal particles. Attached to the corona, and departing from it, there is a jet, where non-thermal particles are strongly accelerated and emit predominantly through optically thin synchrotron. The dominant emission process in the jet scenario can vary depending on the nature of the accreting object and its specific spectral state.  \cite{markoff01}, for example, found that jet synchrotron emission could account for the whole SED of the BH XTE J1118+480 in the hard state.  This means that synchrotron is dominating from the radio through the X-ray. \cite{gallo07} found instead that the X-ray quiescent emission of the BH A0620-00 is primarily produced by inverse Compton processes at the base of the jet. We know that a compact jet \citep[see also][]{millerjones08} is likely driven from the inner region of the accretion disk producing a flat radio spectrum and an excess in the mid-IR.  In corona/Compton models, that can or cannot include the presence of a jet, variability in the mass accretion rate would lead to a change in density in the corona, and hence a change in the probability of Compton scattering in the corona. In the typical unsaturated Comptonization scenario, the slope of the power law in the energy spectrum will depend on the optical depth, and hence the mass accretion rate. We constrained the maximum fractional variability of the spectrum to be lower than $11\%$, and we demonstrated that the spectral shape remains constant within uncertainties as the flux varies. This would favour a scenario in which the X-ray emission is dominated by synchrotron radiation produced in a jet rather than Comptonization.  In the synchrotron scenario one may naively expect less dependence of the spectral slope on the mass accretion rate and more likely a change in the normalization.  This is because the spectral slope is mostly set by plasma physics (e.g. acceleration probability) and while there may be some dependence on the mass accretion rate, it is likely weak. 

Summarizing, a significant change in the slope of the power law (and so a spectral shape change with the flux) would naively favor a Comptonization dominating scenario, while if the slope is instead steady, but the normalization is clearly changing with mass accretion rate, as we more likely observe here, then the synchrotron dominating scenario would be favoured. This is in agreement with the results of \cite{hynes09}, who found evidence of the presence of a jet in quiescent radio, and IR data. We stress that we are only proposing a possible scenario, and that this must not be taken as a firm conclusion, since there is still a lot of uncertainty about the corona/jet geometry and its microphysics.  However, we emphasize that future simultaneous multiwavelength observations, including the whole source SED, and in particular covering the hard X-ray range, like in the case of the recently launched {\it NuSTAR} satellite, could certainly help in solving the mystery about the physical mechanism generating hard X-ray photons: Compton, synchrotron, and bremsstrahlung emission are expected to show significant differences in the spectrum above 10 keV. 

\subsection{A comparison with Cen X-4}
\label{subsub:cenx4}

In this section we compare V404 Cyg with Cen X-4, likely the best studied quiescent NS LMXB, that we recently monitored with a similar campaign \citep{bernardini13}. Despite their intrinsic difference, due to the nature and the mass of the compact object, and to the length of the binary orbital period ($\sim15$ hours for Cen X-4, and $\sim6.4$ days for V404 Cyg)  the two sources share several common properties. They both show a strongly variable X-ray light curve, with a multi-peaked, flare-like structure, with variability on short timescales ($\sim$days).  They are both variable over a wide range of timescales: from hundreds of seconds and tens of minutes for Cen X-4 and V404 Cyg respectively to years. However, in the case of Cen X-4 the X-ray light curve is very likely generated by a red noise power spectrum, while the X-ray light curve of V404 Cyg could be generated by either a flicker noise power spectrum or by a white noise power spectrum. Consequently, we can not determine if it is due to a correlated process or not. Both systems showed a clear correlation between their X-ray and optical emission \citep[see][for the case of V404 Cyg and for the case of Cen X-4 respectively]{hynes04,bernardini13}. The correlation was interpreted in either case as X-ray reprocessing (from the companion star and the accretion disk in the case of Cen X-4 and V404 Cyg respectively). We note that the X-ray emission of Cen X-4 is also strongly correlated with that in the UV band, however, no simultaneous X-ray/UV coverage was available for V404 Cyg. For both sources the X-ray spectral shape remained approximately constant as the X-ray flux varied during quiescence. All these similarities likely imply, as suggested by \cite{cackett13,bernardini13}, that a low level of accretion is still occurring for Cen X-4 also in quiescence, as it occurs for V404 Cyg.

Comparing the power-law index between the two sources, however, we see a difference. The power law spectral index of Cen X-4 is slightly harder 1.4--2.0 compared to 2.10--2.35 of V404 Cyg. The two different spectral shapes could indicate two distinct emission mechanisms at work. BH LMXBs could be in a jet dominated state also in quiescence, like in the case of A0620-00 \citep{gallo07},  where the jet is dominating from the radio (through synchrotron emission) through the soft X-ray (through Compton emission)  and also in the case of V404 Cyg itself \citep[][]{hynes09}, where a steady jet is  producing synchrotron emission in the radio and likely mid-IR bands (and it might also somehow influence the X-ray emission). On the contrary, radio emission has never been detected from NS LMXBs in quiescence.  Recent studies suggest instead that NSs could never reach a jet dominated state, always remaining X-ray dominated objects \citep{migliari06}. More recently, \cite{bernardini13} suggested that the power law of Cen X-4 could be produced by residual accretion, since it is strictly linked to the thermal component arising from the NS surface, they are both changing in tandem as the flux vary. Moreover, for NSs in general, when their spectrum is dominated by the non thermal power law component (that is not the case of Cen X-4) the presence of the NS magnetic field could also provide further mechanisms,  different from a jet, for creating a power-law like spectrum. For example residual accretion on the NS magnetosphere could occur, or alternatively a shock could form due to the interaction of the wind of a quenched pulsar and the accreting matter \citep[][respectively]{arons93,campana98}.
 
\section{Summary}
\label{sec:summary}

We analysed 34 \Sw\ X-ray observations of the BH LMXB V404 Cyg in quiescence, with the goal of characterizing the source variability. The main results of this work are briefly summarized in the following.

The X-ray light curve is strongly variable, showing several flares with count rate changes of a factor of $\sim5-8$ on timescale of hours. We measure the root mean square variability F$_{\rm var}=57.0\pm3.2\%$.

The first order structure function is consistent with both a flicker noise power spectrum (of index -1) and a white noise power spectrum (of index 0). This implies that the source is variable at all the timescales inspected here (days-months). This completes the wide range of variability timescales already know for the source: from tens of minutes to years.

We produced three averaged X-ray spectra as a function of the source count rate, low $<0.01$ c/s, medium 0.011--0.020 c/s, and high $>0.020$ c/s. We found the spectrum to be well fitted by an absorbed power-law model. We measure a power law index $\Gamma=2.10-2.35$ and found that it is constant within$ \sim1\sigma$, as the flux changes (also between single two-day spaced observations). We found that the maximum fractional variability of the energy spectrum slope is $\Delta\Gamma<11\%$.

Concerning the nature of the matter in the inner accretion flow, our results could naively favour the presence of a jet (synchrotron dominated) as in such a scenario changes in the X-ray luminosity would likely produce only a change in overall flux and not a significant change in spectral shape in the X-ray band.

\section*{Acknowledgments}
FB thanks Michele Donato and Jean-Charles Tropato for their useful help about coding. We thank Sera Markoff and Elena Gallo for useful discussions regarding jet models in quiescence. This work made use of data supplied by the UK Swift Science Data Center at the University of Leicester. This research has made use of the XRT Data Analysis Software (XRTDAS) developed under the responsibility of the ASI Science Data Center (ASDC), Italy.
  
\bibliographystyle{mn2e}
\bibliography{biblio}

\vfill\eject

\end{document}